\documentclass[]{aastex631}

\newcommand\aastex{AAS\TeX}

\received{October 20, 2023}
\revised{December 30, 2023}
\accepted{January 8, 2024}

\submitjournal{ApJ}
\usepackage{hyperref}

\shorttitle{Filament model of solar-type star}

\usepackage{comment}
\usepackage{amsmath}
\usepackage{url}

\usepackage{ulem}

\graphicspath{{./}{figs/}}

\begin{document}

\shorttitle{\aastex\ draft}
\shortauthors{Ikuta and Shibata}

\title{Simple Model for Temporal Variations of H$\alpha$ Spectrum \\ by an Eruptive Filament from a Superflare on a Solar-type Star}

\correspondingauthor{Kai Ikuta}
\email{kaiikuta@g.ecc.u-tokyo.ac.jp}

\author[0000-0002-5978-057X]{Kai Ikuta}
\affiliation{Department of Multidisciplinary Sciences, The University of Tokyo, 3-8-1 Komaba, Meguro, Tokyo 153-8902, Japan}

\author[0000-0003-1206-7889]{Kazunari Shibata}
\affiliation{Kwasan Observatory, Kyoto University, 17 Ohmine-cho, Kita-Kazan, Yamashina, Kyoto 607-8471, Japan}
\affiliation{Department of Environmental Systems Science, Doshisha University, 1-3 Tataramiyakodani, Kyotanabe, Kyoto 610-0394, Japan}

\begin{abstract}
Flares are intense explosions on the solar and stellar surfaces, and solar flares are sometimes accompanied by filament or prominence eruptions. Recently, a large filament eruption associated with a superflare on a solar-type star EK Dra was discovered for the first time. The absorption of the H$\alpha$ spectrum initially exhibited a blueshift with the velocity of $510$ (km s$^{-1}$), and decelerated in time probably due to gravity. Stellar coronal mass ejections (CMEs) were thought to occur, although the filament eruption did not exceed the escape velocity under the surface gravity. To investigate how such filament eruption occur and whether CMEs are associated with the filament eruption or not, we perform one-dimensional hydrodynamic simulation of the flow along an expanding magnetic loop emulating a filament eruption under adiabatic and unsteady conditions. The loop configuration and expanding velocity normal to the loop are specified in the configuration parameters, and we calculate the line-of-sight velocity of the filament eruption using the velocities along and normal to the loop. We found that (i) the temporal variations of the H$\alpha$ spectrum for EK Dra can be explained by falling filament eruption in the loop with longer time and larger spatial scales than that of the Sun, and (ii) the stellar CMEs are also thought to be associated with the filament eruption from the superflare on EK Dra, because the rarefied loop unobserved in the H$\alpha$ spectrum needs to expand faster than the escape velocity, whereas the observed filament eruption does not exceed the escape velocity.
\end{abstract}


\section{Introduction} \label{ref:intro}
Solar and stellar flares are intense explosions in the solar and stellar atmosphere, and have been observed from radio to X-rays \citep[for reviews,][]{Shibata11,Benz17}. Solar flares are sometimes accompanied by filament or prominence eruptions observed respectively in emission or absorption of H$\alpha$ spectrum \citep[for a review,][]{Parenti14}, and the eruptions subsequently could induce coronal mass ejections (CMEs) \citep[for reviews,][]{Chen11, Webb12, Cliver22} if their velocity is sufficiently larger than the solar escape velocity \citep[e.g.,][]{Gopalswamy03}.
It has been reported that solar-type stars (G-dwarfs) cause superflares
with ten to ten hundred times larger energy than that of largest solar flares \citep{Maehara12,Shibayama13,Notsu19,Feinstein20,Okamoto21,Tu21,Jackman21,Namekata22,Namekata_superflare, Pietras22,Yamashita22}.
Many spectroscopic and multiwavelength observations of stellar flares have been performed to investigate the radiation mechanism and mass ejections associated with flares \citep[for a review,][]{Leitzinger22}. Blueshifted spectrum lines have been widely observed on M-dwarfs during stellar flares \citep[e.g.,][]{Houdebine90, Eason92, Gunn94, Crespo06,Hawley07}. In particular, the temporal variation of a blueshifted H$\alpha$ spectrum associated with stellar flares has also been reported as a signature of mass ejections from the star as M-dwarfs \citep[][]{Fuhrmeister08, Fuhrmeister11, Vida16,Honda18,Fuhrmeister18, Vida19,Muheki20, Muheki_evlac, Maehara21,Koller21,Notsu23}, K-dwarfs \citep[][H.Maehara et al., in preparation]{Flores17}, a G-dwarf \citep[][]{Namekata22,Namekata23}, and an RS CVn-type star \citep{Inoue23}. It has also been discussed that the blueshifted H$\alpha$ spectrum observed in the early phase of solar flares \citep[][]{Svestka62} could be a result of the cool upflow \citep[e.g.,][]{Canfield90,Tei18} or an absorption by the cool downflow \citep[e.g.,][]{Heinzel94}, both of which are associated with chromospheric evaporation.

\citet{Namekata22} reported that a temporal enhancement of blueshifted
 absorptions on the H$\alpha$ spectrum was associated with a superflare as a signature of a large filament eruption on an active solar-type star EK Draconis (EK Dra; spectral type of G1.5V) \citep[][]{Strassmeier98}. 
The absorption of H$\alpha$ spectrum initially exhibited the blueshift with the velocity of $510$ (km s$^{-1}$), and decelerate probably with the gravity to the redshift with the velocity of $200$ (km s$^{-1}$). The longer temporal variation of 
the H$\alpha$ absorption for EK Dra can be explained as the filament eruption with larger spatial scale than that for the Sun through the spatially integrated spectrum as a Sun-as-a-star analysis \citep[e.g.,][]{Namekata_sun, Otsu22}.
Large stellar CME was thought to be associated with the large filament eruption because a CME was also associated with the solar filament eruption \citep{Seki21}.
On the other hand, it has been suspected that the stellar CME was not associated because the large filament eruption in this case did not exceed the escape velocity of $670$ (km s$^{-1}$).
Based only on the observation of H$\alpha$ spectrum, we can not understand how such a large filament eruption falls and whether stellar CME occurs or not in association with the filament eruption with smaller velocity than the escape velocity.

In this study, we examine the temporal variation of the H$\alpha$ spectrum with a simple model for the filament eruptions on the Sun and EK Dra. We perform hydrodynamic simulation emulating a filament eruption in an expanding magnetic loop with time. The configuration of the magnetic loop and expanding velocity normal to the loop are specified in the configuration parameters, and we solve the one-dimensional hydrodynamic equations of the flow along the expanding loop under adiabatic and unsteady conditions. We also calculate the line-of-sight velocity of the filament from the velocities along and normal to the loop. The purposes of this paper are (i) comparing the simple model with observations of the filament eruptions on the Sun and EK Dra, (ii) clarifying which parts of the loop emanate in H$\alpha$ spectrum as the filament eruption, and (iii) predicting whether CMEs were associated with the large filament eruption on EK Dra beyond the escape velocity. The remainder of this paper is organized as follows.
Basic equations, numerical setups, and observations of the filament eruption, are described in Section \ref{ref:method}. The result and discussion of hydrodynamics, modeled H$\alpha$ spectrum, and comparison with solar H$\alpha$ image are described in Section \ref{sec:result}.
The conclusion and future prospects of our model are described in Section \ref{sec:conclusion}. The case under an optically thick condition is also discussed in Appendix \ref{sec:thick}.

\section{Method} \label{ref:method}
\subsection{Basic equations} \label{sec:fluid}
\subsubsection{Curvilinear coordinate}
\begin{figure*}[p]   
\plotone{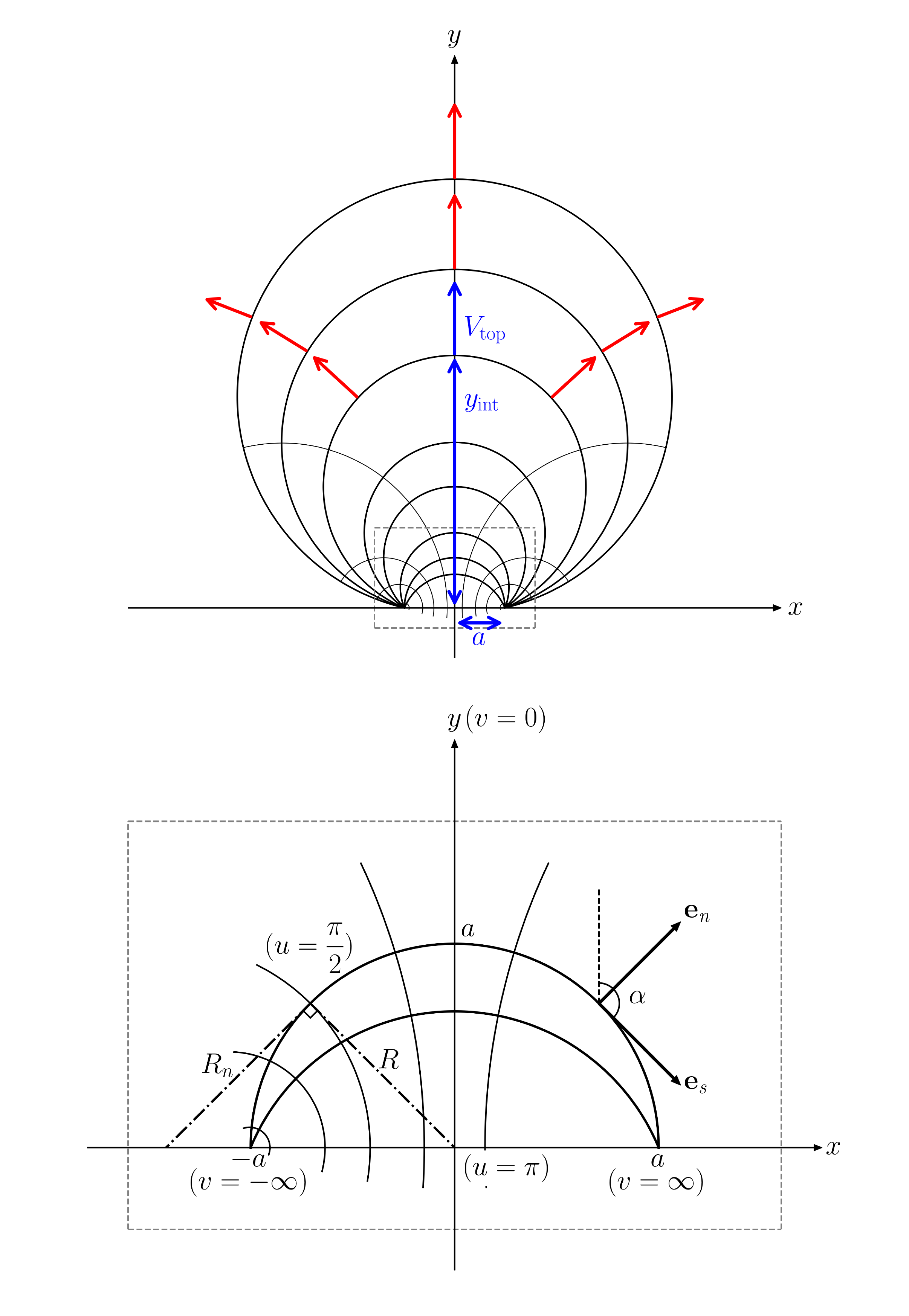} 
\caption{(Top) Configuration of an expanding loop with time (red) as the bipolar coordinate $(v,u)$ in the Cartesian coordinate $(x,y)$. The configuration parameters are described for the half distance of the two line poles at the loop bottom $a$, initial height of the loop top $y_{\rm int}$, and initial normal velocity of the loop top $V_{\rm top}$ (blue).
(Bottom) Enlarged description of the bipolar coordinate $(v,u)$ and curvilinear coordinate $(s,n)$ in the Cartesian coordinate $(x,y)$ \citep[Fig.2 in][]{Shibata80}.} \label{fig:coord}
\end{figure*}

We solve one-dimensional hydrodynamic equations along an expanding magnetic loop with time (Figure \ref{fig:coord}). 
The equations of a flow in the moving magnetic flux tube are originally formulated from 
two-dimensional equations in the curvilinear coordinate $(s,n)$ determined by the configuration of arbitrary magnetic field \citep{Kopp76,Venka81}, which represent the coordinate along and normal to the tube, respectively.
We derive the energy equation along the loop for the adiabatic case instead of the equation of state for the isothermal case \citep{Kopp76} and polytropic case \citep{Venka81}.
In particular, the curvilinear coordinate $(s,n)$ can be transformed to other curvilinear coordinate $(v,u)$ with the scale factor $h$ in the form of 
\begin{align} \label{eq:scale_factor}
ds &=h^{-1}dv \notag \\ 
dn &= -h^{-1} du.
\end{align}
Then, in the coordinate along the loop $v$ from the restframe of the coordinate normal to the loop $u(t)$, the continuity equation, equation of motion, and energy equation are represented by

\begin{equation} \label{eq:continuity}
\frac{D}{Dt} (\rho A) + \frac{D}{Dv} (\rho V_s A h) -\frac{\rho V_s A}{R_n} + \frac{\rho V_n A}{R} = 0,
\end{equation}

\begin{align} \label{eq:motion}
\frac{D}{Dt} (\rho V_s A) &+ \frac{D}{Dv} \{(\rho V^2_s +p ) A h\} -\frac{\rho V^2_s A}{R_n} \notag \\ &+\frac{2\rho V_s V_n A}{R} +\frac{\rho V^2_n A}{R_n} 
+ \rho A h \frac{D \phi}{Dv} = 0,
\end{align}
and
\begin{align} \label{eq:energy}
\frac{D}{Dt} \{ (\frac{1}{2}\rho V_s^2 &+\frac{p}{\gamma-1}) A \} + \frac{D}{Dv}\{(\frac{1}{2}\rho V_s^2+\frac{\gamma p}{\gamma-1}) V_s A h\} \notag \\ &- \frac{\rho V_s^3  A}{2R_n}
+  \frac{3\rho V_s^2 V_n A }{2R} 
+\frac{\rho V_s  V_n^2 A}{R_n} \notag \\ 
 &- \frac{\gamma p V_s A}{(\gamma-1) R_n}  + \frac{(2\gamma-1) p V_n A}{(\gamma-1) R}
 \notag \\  &+ p  A \frac{D}{Dt} \ln f(u) +\rho V_s A h \frac{D \phi}{Dv} =0,
\end{align}
where $\rho$, $p$, and $V_{s}$ are the density, pressure, and velocity along the loop, respectively.
The parameters $A$, $V_n$, $R$, $R_n$, $\phi$ are the cross-sectional area, velocity normal to the loop, the curvature of the loop in $v$, that in $u$, and gravitational potential, respectively.
The derivatives are represented by
\begin{align}
\frac{D}{Dt} &= \frac{\partial}{\partial t} + V_n \frac{\partial}{\partial n} \notag \\ 
\frac{D}{Dv} &= \frac{\partial}{\partial v}.
\end{align}
The cross-sectional area $A = h^{-1} f(u)$ is a free parameter tuning the normal velocity $V_n$ as the following equation:
\begin{equation} \label{eq:cross}
\frac{\partial V_n}{\partial n} = \frac{D}{Dt}  \ln A,
\end{equation}
which is derived from the induction equation along the loop
\begin{equation} \label{eq:induction}
\frac{\partial B}{\partial t} = -\frac{\partial}{\partial n} (V_n B),
\end{equation}
and the divergence-free condition of magnetic field $BA$ = const., where $B$ is the magnetic field strength along the loop.

We employ the simple representation $f(u)=\sin ^ 2 u$ so that the cross-section at the loop top expands and the normal velocity at the loop top moderately accelerates over time. 
We note that the induction equation normal to the loop $ \partial (V_n B)/ \partial s = 0$ is also satisfied. The cross-sectional area $A$ can be multiplied by a constant, and we select the constant to one for simplicity. In this study, we also set an unit length in the direction perpendicular to the $(s,n)$-plane so that $A$ has the dimension of the area (cm$^2$).
The last terms as the derivative of $\phi$ in Equations \ref{eq:motion} and \ref{eq:energy} are contributions from the gravitational potential in the form of the Cartesian coordinate $(x,y)$:
\begin{equation}
\phi (v,u) = - \frac{G M_{\sun}}{\{x^2+(y+b)^2\}^{1/2}}.
\end{equation}
The direction of the gravity is toward the center of the star $(0,-b)$ and $b=\sqrt{R_{\sun}^2-a^2}$.
$G=6.6743 \times 10^{-11}$ (m$^3$ kg$^{-1}$ s$^{-2}$), $M_{\sun}=1.988 \times 10^{30}$ (kg), and  $R_{\sun}=6.957 \times 10^8$ (m) are the gravitational constant, solar mass, and solar radius, respectively. The solar surface gravity is calculated as $g_{\sun}=G M_{\sun} / R_{\sun}^2 = 2.741 \times 10^2$ (m s$^{-2}$).
For simplicity, we adopt solar mass and radius in the case of EK Dra because various values near the solar one are reported as the mass and radius of EK Dra \citep{Waite17,Senavci21}. 

\subsubsection{Bipolar coordinates}
Observations of solar CMEs
and filament eruptions often show self-similar expansion of magnetic loops in which magnetic field lines have circular configuration
as illustrated in \cite{Aschwanden17} (see also \citet{Dere97} for observation
of typical self-similar expansion of CMEs, 
\citet{low84}, and \citet{Gibson98}, for theory of self-similar expansion)\footnote{RHESSI Nuggets: \url{https://sprg.ssl.berkeley.edu/~tohban/wiki/index.php/Bridging_solar_flares_to_coronal_mass_ejections}}.
To approximately reproduce the self-similarly expanding loop of CMEs and filament eruptions, we assume the configuration of bipolar magnetic field by two line currents (Figure \ref{fig:coord}) and introduce bipolar coordinate defined by this bipolar magnetic configuration as in \cite{Shibata80}.
Then, the scale factor of the bipolar coordinate is given by
\begin{equation}
h=\frac{\cosh v - \cos u}{a}
\end{equation}
from the curvilinear coordinate in the form of Equation \ref{eq:scale_factor}, and $a$ is the location of the foci in the Cartesian coordinate $(x,y) = (\pm a, 0)$. The Cartesian coordinate is transformed from the bipolar coordinate as 
\begin{align} \label{eq:cartesian}
x&=\frac{\sinh v}{h} \notag \\
y&=\frac{\sin u}{h}.
\end{align}
The curvature of the loop in $v$ and $u$ are given by
\begin{align}
R &= \frac{a}{\sin u} \notag \\ 
R_n &= \frac{a}{\sinh v},
\end{align}
respectively.
The length from $v=0$ (loop top) in the curvilinear coordinate is integrated from Equation \ref{eq:scale_factor} as 
\begin{equation} \label{eq:scale}
    s = \frac{2a}{\sin u} \tan^{-1} [\frac{(1+\cos u) \tanh (v/2)}{\sin u}].
\end{equation}

\subsection{Numerical setups} \label{sec:setup}
\subsubsection{Initial and boundary conditions} \label{sec:condition}
We inflate the magnetic loop emulating a filament eruption partly in it with time.
The range of the coordinate $v$ is set from $v=0$ to $3$, and
we adopt the symmetric and free boundary at the loop top ($v=0$) and bottom ($v=3$), respectively. We set the mesh of $dv \propto h (v, u(t=0)\equiv u_0)$ as $ds={\rm const.}$ at the initial time and use $1.2 \times 10^5$ meshes to resolve the loop sufficiently.

As described in Figure \ref{fig:coord}, the configuration of the loop is specified in three parameters: 
\begin{itemize}
 \item[(1)] the half distance between the two line poles $a$ (km), which corresponds to the size of an active region at the loop bottom
 \item[(2)] the initial height of the loop top $y(v=0,u=u_0)=y_{\rm int}$ (km)
 \item[(3)] the initial normal velocity of the loop top $V_n(v=0,u=u_0) = V_{\rm top}$ (km s$^{-1}$).
\end{itemize}
These parameters are determined so that the motion of mass reproduce the observed temporal variation of H$\alpha$ spectrum for the Sun and EK Dra.
Then, $u_0 = u(t=0)$ is also given by Equation \ref{eq:cartesian} as the initial value of $u(t)$:
\begin{equation}
u_0 = \cos^{-1}\frac{y_{\rm int}^2-a^2}{y_{\rm int}^2+a^2}.
\end{equation}

We also assume that there is a cool and dense filament with the uniform temperature $T=10^4$ K and density $\rho (v,u_0) = \rho_0 = 10^{-13}$ (g cm$^{-3}$) in a localized region $v$ between $v_{\rm min}$ and $v_{\rm max}$ ($0<v_{\rm min}<v<v_{\rm max} < 3$) at the initial magnetic loop ($u=u_0$).
Then, the uniform pressure is also given by $p (v,u_0) = p_0 = \rho R_g T/ \mu = 0.166$ (dyn cm$^{-2}$), where $R_g=8.31 \times 10^7$ (erg K$^{-1}$ mol$^{-1}$) and $\mu=0.5$ (g mol$^{-1}$) are the gas constant and mean molecular weight, respectively.
For the outside of the filament ($0<v < v_{\rm min}$, $v_{\rm max}< v < 3$), we assume that there is a hot and rarefied corona with the uniform temperature $T=10^6$ K, density $\rho (v,u_0) = 0.01\rho_0 = 10^{-15}$ (g cm$^{-3}$), and pressure $p (v,u_0) = p_0 = 0.166$ (dyn cm$^{-2}$) so that the coronal plasma is initially in a pressure balance with the filament plasma.
We assume further that the initial velocity along the loop is in the rest $V_s (v,u_0) = 0$ (km s$^{-1}$) throughout the loop.
These initial temperature, density, pressure, and velocity along the loop are fixed for all adopted models in this study.

Under these parameters and conditions, we numerically solve one-dimensional hydrodynamic equations (Equations \ref{eq:continuity}, \ref{eq:motion}, and \ref{eq:energy}) using modified Lax-Wendroff scheme \citep{Rubin67} with an artificial viscosity. Then, we can obtain the motion of mass along the expanding loop as the density $\rho$, pressure $p$, and velocity along the loop $V_s$. The temperature $T$ is also calculated from the equation of state.

\subsubsection{Calculations of the line-of-sight velocity} \label{sec:vobs}

\begin{figure*}[p]   
\plotone{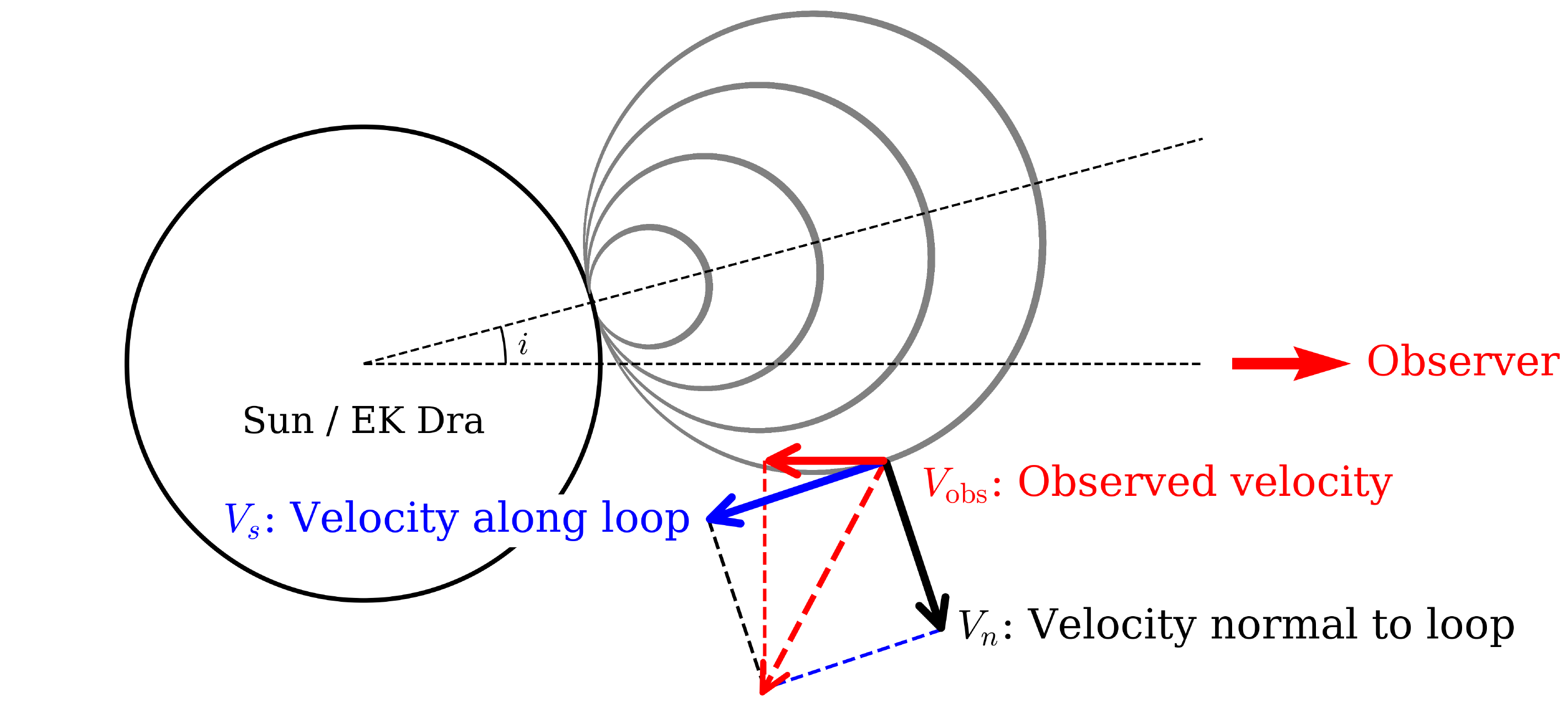} 
\caption{The observed velocity $V_{\rm obs}$ (red) is calculated by projecting the composite velocity of the velocity along the loop $V_s$ (blue) and velocity normal to the loop $V_n$ (black) in the direction to the line of sight (red) from the Equations \ref{eq:los} and \ref{eq:vxy}.} \label{fig:los}
\end{figure*}

As in Figure \ref{fig:los}, the velocity $V_{\rm obs}$ in the direction of the line of sight corresponds to the observed velocity in the H$\alpha$ spectrum:
\begin{equation} \label{eq:los}
V_{\rm obs}= -V_x \sin i - V_y \cos i,
\end{equation}
where $i$ is the viewing angle between the line of sight and $y$-direction. $V_x$ and $V_y$ are calculated by a composite of velocity $V_s$ along the loop and velocity $V_n$ normal to the loop: 

\begin{align} \label{eq:vxy}
V_x &=  V_s \sin \alpha - V_n \cos \alpha, \notag \\
V_y &= V_s \cos \alpha + V_n \sin \alpha , \notag \\
\tan \alpha &= - \frac{1- \cosh v \cos u}{\sinh v \sin u},
\end{align}

where $\alpha$ is the angle between $y$- and $s$-directions (Figure \ref{fig:coord}). Instead of solving the radiative transfer equations to calculate H$\alpha$ spectrum, we simply assume that the erupting filament is optically thin and the H$\alpha$ intensity is in proportion to the mass to the direction of the line of sight by considering the observed velocity $V_{\rm obs}$.
Thus, although the assumption in this study is very idealized, even such a simple model would be useful for comparing the model of an erupting filament with observations, as discussed in Section \ref{sec:result}. 

The temporal variation of the modeled H$\alpha$ spectrum is calculated by summing up the mass $\rho_j ds_j dn_j = \rho_j A_j h_j^{-1} dv_j$ with each observed velocity $V_{{\rm obs},j}$ for $j$-th mesh point along the loop $v$ and on the disk ($|x \cos i - (y+b) \sin i|\leq R_{\sun}$) for each time of $t$.
In addition, we assume that the mass with the temperature $3 \times 10^3 \leq T \leq 3 \times 10^4$ K contribute to the variation of the modeled H$\alpha$ spectrum because the mass with $T \simeq 10^4$ K is typically observed in the H$\alpha$ spectrum.
Assuming symmetric initial condition for the mass distribution of the filament with respect to the center of the loop ($v=0$), we also include the mass with the coordinate from $v=-3$ to $v=0$ ($x \rightarrow -x$) because the velocity to the line of sight is different from those from $v=0$ to $v=3$ under the viewing angle of $i \neq 0^{\circ}$.
In particular, to compare with the normalized intensity of the observed H$\alpha$ spectrum clearly, we normalize the mass with each velocity by its maximum in the temporal variation of the modeled H$\alpha$ spectrum and scale it by the normalized intensity of the observed H$\alpha$ spectrum.

\subsubsection{Survey for parameters and viewing angle} \label{sec:parameter_survey}
First, we perform the parameter survey to investigate whether the maximum velocity and deceleration timescale by the gravity can correspond to the temporal variation of the observed H$\alpha$ spectrum on the Sun and EK Dra, as discussed in detail in Section \ref{sec:result}. 
As a result, suitable parameter combinations to reproduce the observed H$\alpha$ spectrum are found to be $(a, y_{\rm int}, V_{\rm top}) = (1.0 \times 10^4 \textrm{ km}, 4.5 \times 10^4 \textrm{ km}, 1.5 \times 10^2 \textrm{ km s}^{-1})$ and $(2.5 \times 10^4 \textrm{ km}, 3.5 \times 10^5 \textrm{ km}, 5.5 \times 10^2  \textrm{ km s}^{-1})$ for the Sun and EK Dra, respectively. The parameters are almost determined by the velocity at the blueshift of the H$\alpha$ spectrum and the falling timescale of the mass.

Second, the particles with the different initial locations in the loop are traced using the parameter combination encompassing the observed H$\alpha$ spectrum, and we search for the initial location of the density in the range of $v_{\rm min} < v < v_{\rm max}$ to reproduce the observed H$\alpha$ spectrum under the viewing angle $i=0^{\circ}$ (i.e., $V_{\rm obs} = - V_y$).
However, in the case of EK Dra, the large filament eruption is out of the stellar disk from the middle of its fall because the expanding loop becomes larger than the stellar disk. Thus, we also search for the viewing angle $i$ for EK Dra so that the filament eruption is in the disk.
As a result, we found that $(v_{\rm min},v_{\rm max})$ = $(0.11,0.20)$ and $(0.055,0.105)$, corresponding to the lengths from the loop top ($s(v_{\rm min},u_0)$ and $s(v_{\rm max},u_0)$) = ($1.14 \times 10^4$, $1.99 \times 10^4$) and ($1.29 \times 10^5$, $2.23 \times 10^5$) (km), for the Sun and EK Dra, respectively. The suitable viewing angle for EK Dra is also found to be $i=40^{\circ}$ (deg).
In addition, to suppress the numerical diffusion due to the discontinuity of the initial density, we smoothly connect the densities of filament $(=\rho_0)$ and corona $(=0.01\rho_0)$ by the sigmoid function: $\rho (v,u_0) = 0.5 \rho_0 \tanh \{ 500 (v-v_{\rm min})  \} +0.505$ or $0.5 \rho_0 \tanh \{ 500(v_{\rm max}-v) \} +0.505$ for $v \leq (v_{\rm min}+v_{\rm max})/2$ or $(v_{\rm min}+v_{\rm max})/2 < v$, respectively.

\subsection{Observations of the filament eruption} \label{sec:obs}

To compare the model with observed temporal variations of H$\alpha$ spectrum reported in \cite{Namekata22}, we briefly describe the observations of the filament eruption on the Sun and EK Dra.

The solar filament eruption was associated with a C5.1-class solar flare observed from 07:52:09 UT on 7 July 2016 by the Solar Dynamics Doppler Imager mounted on the Solar Magnetic Activity Research Telescope \citep[SMART/SDDI;][]{Ichimoto17}  at Hida Observatory in Kyoto University.
The temporal variation of the H$\alpha$ spectrum (centered on 6562.8 ${\rm \AA}$) from the pre-flare level is calculated as the part of flare region relative to the quiet photosphere \citep[for details, Fig.2 and Extended Data Fig.4 in][]{Namekata22}. The filament erupted until 10 (min) with the blueshift, and after that fall to the solar surface with the redshift. Solar CMEs were also associated with the filament eruption \citep{Seki21}.

On the solar-type star EK Dra, a superflare with the bolometric energy of $2.0 \times 10^{33}$ (erg) was observed by the 3.8m \texttt{Seimei} telescope \citep{2020PASJ...72...48K} at Okayama Observatory in Kyoto University and the 2m \texttt{Nayuta} telescope at Nishi-Harima Astronomical Observatory in Hyogo University on 5 April 2020 \citep[for details, Fig.1 in][]{Namekata22}.
The blueshifted absorption of H$\alpha$ spectrum from the pre-flare level was associated with the superflare 25 minutes later than the peak time (BJD 2458945.2). The blueshifted absorption on H$\alpha$ spectrum has been observed until 70 (min) as a signature of a large filament eruption with the blueshift, and after that fell to the stellar surface with the redshift. The temporal variation of the H$\alpha$ spectrum is similar to the solar one with different time and spatial scales, and stellar CMEs were thought to be also associated with the large filament eruption.

\section{Results and Discussion} \label{sec:result}

\subsection{The Sun}

\subsubsection{Hydrodynamics} \label{sec:hydrosun}
\begin{figure*}[p]   
\plotone{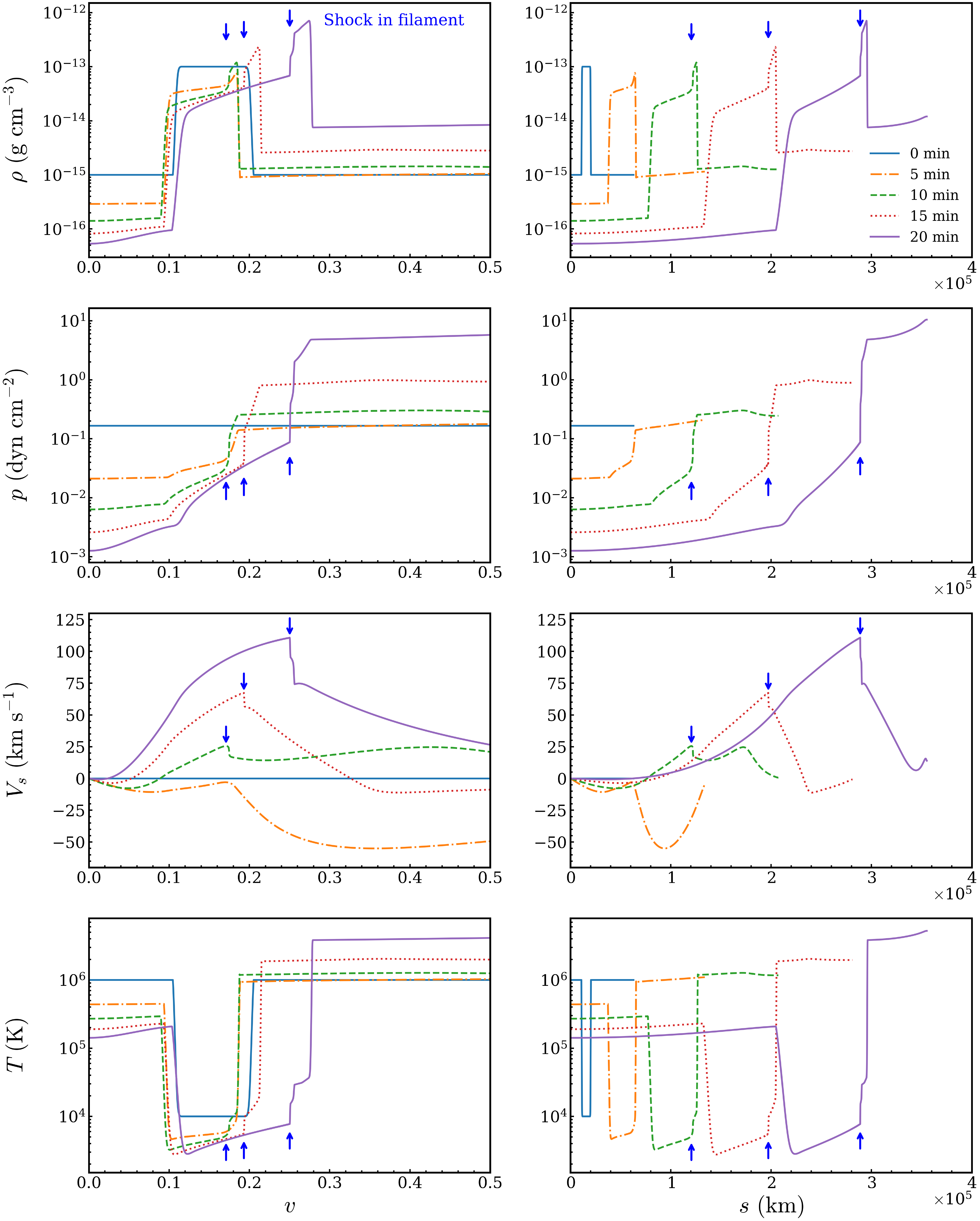} 
\caption{The temporal variation of density $\rho$, pressure $p$, velocity $V_s$, and temperature $T$, with the coordinate $v$ (left) and $s$ (right) for the Sun at the time of $t=$ 0, 5, 10, 15, and 20 (min).
The shock wave generated inside the filament is indicated for $t=$ 10, 15, and 20 (min) (blue arrow). The range of $0 \leq v \leq 0.5$ is only exhibited to enlarge the part of the filament eruption for clarity.
The length from the loop top $s$ can be transformed from the coordinate $(v,u)$ with Equation \ref{eq:scale}.} \label{fig:fluid_sun}
\end{figure*}

Figure \ref{fig:fluid_sun} shows the results of the hydrodynamic simulations of the flow along the expanding loop for the Sun (the filament eruption associated with the C5.1-class solar flare on 7 July 2016): the temporal variations of density $\rho$, pressure $p$, velocity $V_s$, and temperature $T$, as a function of the coordinate $v$ ($\leq 0.5$) and $s$ at the time of $t$ = 0, 5, 10, 15, and 20 (min).
The cool and dense filament appears to be roughly in free-fall motion due to the gravity toward the loop bottom (rightward in each panel of Figure \ref{fig:fluid_sun}). Thus, the adiabatic expansion of the plasma at the loop top (corona) and inside the filament leads to significant decrease in the density and pressure. 
This is the reason why the apparent large pressure imbalance appears inside the filament for $t$ = 5 (min). However, it should be noted that dynamical equilibrium is roughly satisfied (i.e., pressure gradient force roughly balances with gravity along the loop) both at the loop top and inside the filament for $t$ = 5 (min) because the temperature in the filament is low so that the pressure scale height is very low. We also note that the pressure balance is satisfied between the corona at the loop top and the top of the filament.
It is of interest to note that a shock wave is generated inside the cool filament because the falling velocity (= 26, 67, and 111 km s$^{-1}$ at the shock) exceeds the local sound speed (= 12, 12, and 15 km s$^{-1}$) for $t$ = 10, 15, and 20 (min). This reverse shock propagates to the upward direction in the falling filament (leftward). Thus, as the filament falls, its mass at the bottom goes through the shock, so that it is compressed and heated by the shock, even if the main part of the filament expands and after that both density and temperature decrease with time. It is also seen that the falling filament compresses coronal plasma just below the filament.

\subsubsection{Comparison with observed H$\alpha$ spectrum} \label{sec:solar_model}
\begin{figure*}[p]
\plotone{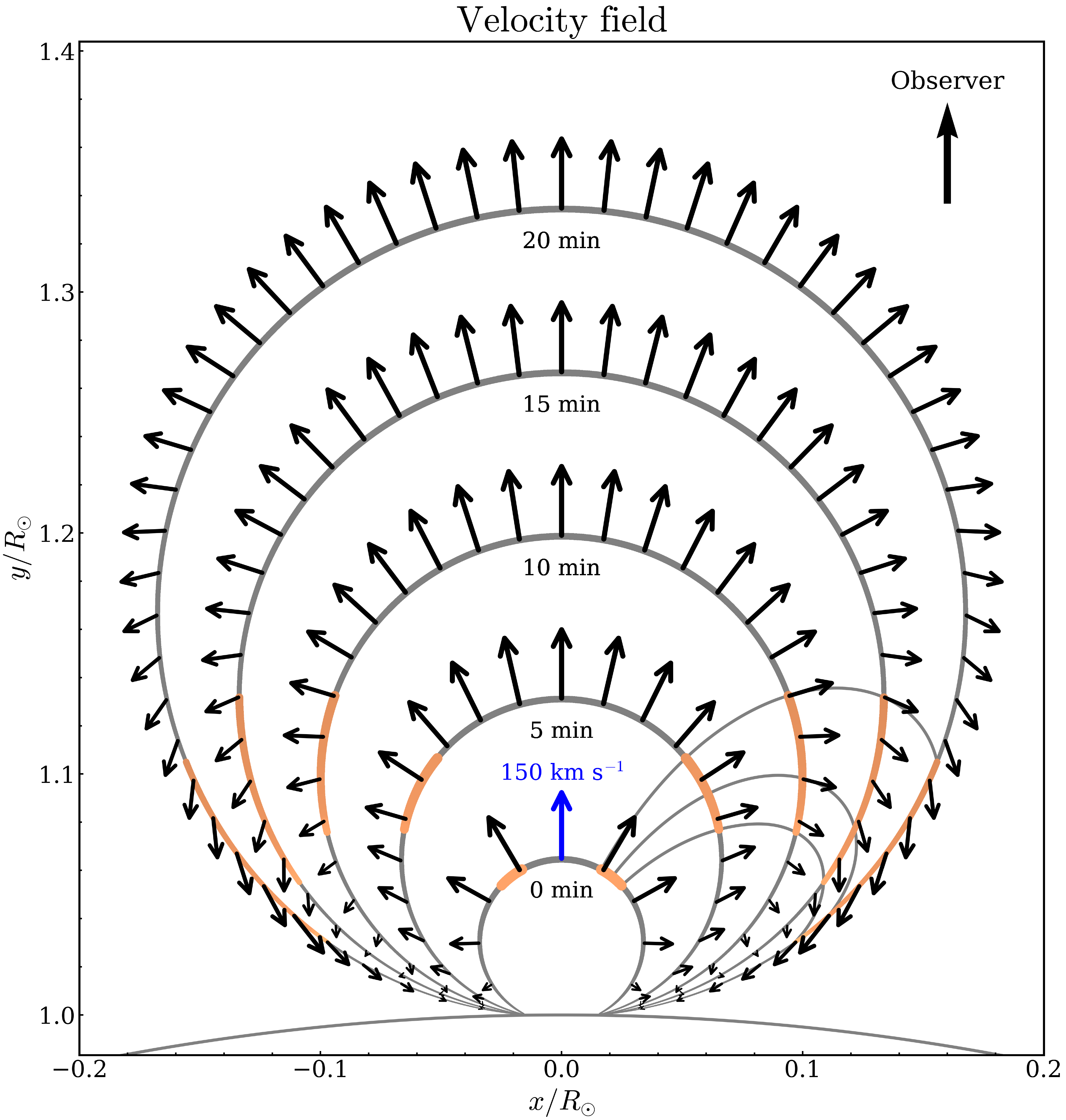}
\caption{Velocity field $(V_x,V_y)$ of the expanding loop in the Cartesian coordinate $(x,y)$ for the Sun (the filament eruption associated with the C5.1-class solar flare on 7 July 2016) at the time of $t=$ 0, 5, 10, 15, and 20 (min). The filament eruption in the loop is colored in orange. The initial normal velocity of the loop top $V_{\rm top}=150$ (km s$^{-1}$) is also colored in blue for comparison of the velocity scale.}\label{fig:vector_sun}
\end{figure*}

\begin{figure*}[p]
\plotone{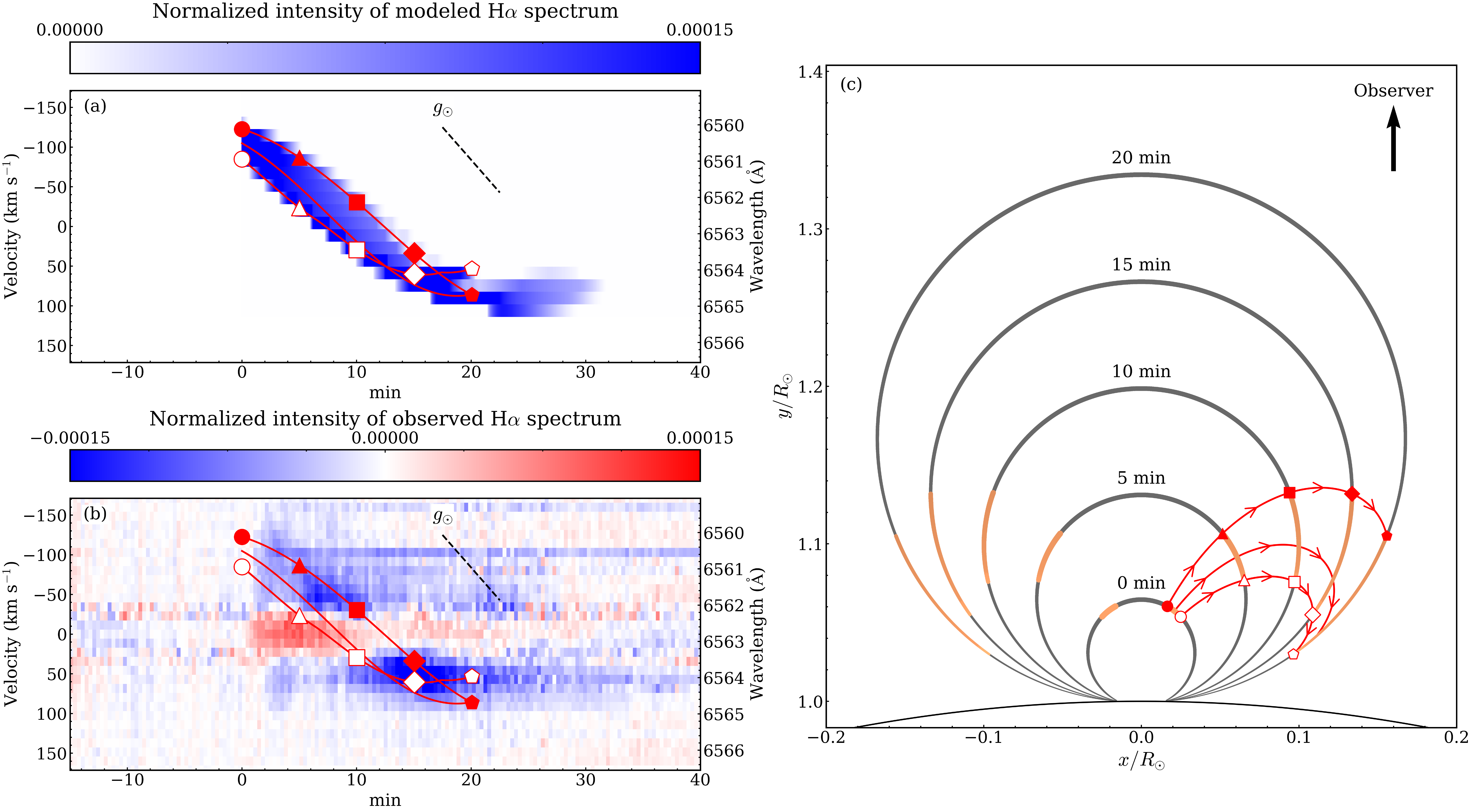}
\caption{(a) Normalized intensity of the modeled H$\alpha$ spectrum (Section \ref{sec:vobs}), (b) normalized intensity of the observed H$\alpha$ spectrum of the filament eruption associated with the C5.1-class solar flare on 7 July 2016 \citep[Fig.2 in][]{Namekata22}, and (c) configuration of the filament eruption in the expanding loop at the time of $t=$ 0, 5, 10, 15, and 20 (min) (orange), for the Sun from the viewing angle of $i=0^{\circ}$.
Red solid lines with each mark show the temporal variations of $V_{\rm obs}$ in (a, b) for the upper, middle, and lower locations of the filament eruption in (c) at each time. The solar surface gravity $g_{\sun}$ is also represented in (a, b) for comparison.
} \label{fig:contour_sun}
\end{figure*}

Figure \ref{fig:vector_sun} shows the result of velocity fields of the expanding loop at the time of $t$ = 0, 5, 10, 15, and 20 (min). The orange parts of the loop correspond to the erupting filament. It is also seen that the filament falls down after $t=$ 10 (min).
Figure \ref{fig:contour_sun} shows (a) the normalized intensity of the modeled H$\alpha$ spectrum reproduced by summing up the normalized mass with each observed velocity in velocity-time diagram under an assumption that the intensity is proportional to the mass (Section \ref{sec:setup}), (b) the normalized intensity of the observed H$\alpha$ spectrum of the solar filament eruption associated with C5.1-class flare on 7 July 2016 (Section \ref{sec:obs}), and (c) configuration of the filament eruption in the loop at the time of $t$ = 0, 5, 10, 15, and 20 (min).
It can be seen that the modeled H$\alpha$ spectrum roughly corresponds to the observed H$\alpha$ spectrum except for the H$\alpha$ emission around the zero velocity. We note that our model does not consider the effect of H$\alpha$ emission from the solar flare.

It should be also noted that the velocity at the loop top is less than the escape velocity even at the initial height: $V_{\rm top} = 150 < \sqrt{2 G M_\sun/(y_{\rm int }+b)}= 599$ (km s$^{-1}$). However, considering the previous observations of comparison between solar filament/prominence eruptions and CMEs \citep{Gopalswamy03, Seki21, Namekata22} as well as self-similar expansion model of CMEs \citep{low84,Gibson98}, we can suggest that our loop containing the erupting filament (as in Figure \ref{fig:vector_sun}) eventually results in CMEs. 
That is, our simple model is based on the continuously expanding model, so that even after the cool and dense filament falls to the loop bottom, the loop top with hot coronal plasma continuously expand to distant radius of the Sun, where the escape velocity is less than the initial velocity of the loop top. In fact, in the case of this particular filament eruption, CMEs were actually observed in association with the filament eruption \citep{Seki21}.

\subsubsection{Comparison with H$\alpha$ image}

\begin{figure*}[p]   
\epsscale{0.8}
\plotone{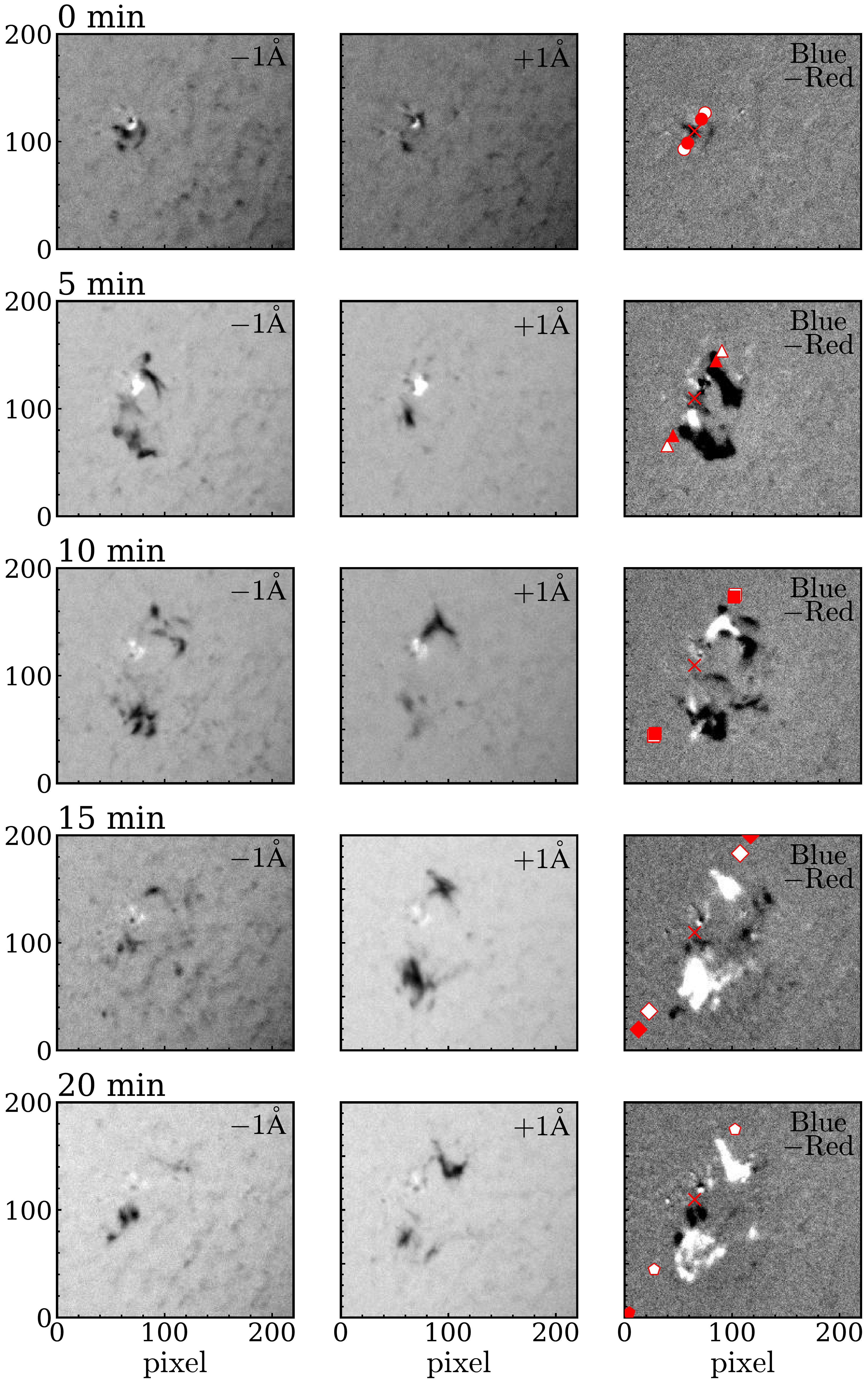} 
\caption{Temporal H$\alpha$ images of the solar filament eruption associated with the C5.1-class solar flare observed by SMART/SDDI on 7 July 2016 at the time of $t=$ 0, 5, 10, 15, and 20 (min) from the start of the flare on 07:52:09 UT \citep[Extended Data Fig.4 in][]{Namekata22}.
The left and middle images are taken at wavelengths of -1 $\mathrm{\AA}$ (-46 km s$^{-1}$) and +1 $\mathrm{\AA}$ (+46 km s$^{-1}$) from the center of the H$\alpha$ spectrum (= 6562.8 $\mathrm{\AA}$). The emission and absorption components are indicated with white and black, respectively.
The right images show the subtraction of the redshift image (+1 to +3 $\mathrm{\AA}$) (white) from the blueshift one (-3 to -1 $\mathrm{\AA}$) (black).
The upper, middle, and lower locations of the filament eruption are marked for each time (Figure \ref{fig:contour_sun}). The center of the loop are also marked in red cross.
We note that 1 (pixel) corresponds to 1.23 (arcsec) = 892 (km).}
\label{fig:compare}
\end{figure*}

The time and spatial scales of the model can be compared to the observed filament eruption since solar filament eruptions can be spatially resolved.
Figure \ref{fig:compare} shows the blueshift H$\alpha$ image ($=6562.8-1{\rm \AA}$), redshift one ($=6562.8+1{\rm \AA}$), and subtracted one of redshift one from blueshift one, for the solar filament eruption reported in \cite{Namekata22} at the time of $t$ = 0, 5, 10, 15, and 20 (min) from the start of the flare, and the locations of the filament eruption in our model are overplotted on each subtracted image. Both of the H$\alpha$ spectrum and image were obtained by the SMART/SDDI (Section \ref{sec:obs}).

Even on the spatially resolved images, the model of the erupting filament, which is marked by red symbols on its top and bottom, roughly correspond to the H$\alpha$ images of the solar filament eruption.
It is shown that the horizontal locations of the filament in the model are slightly shifted from the observed one by several ten percent to a factor of the loop length after 5 (min).
This difference is thought to be due to the simple model by the loop in the bipolar coordinate, and as future works, the model should be updated for the configuration of the magnetic field. However, even in this simple model, this approximate correspondence of the time and spatial scales is an intriguing result for the validity of our model (or future possibility of the model development), considering that a simple model of an expanding loop is adopted for modeling the filament eruption.
Thus, this simple model can be adapted to the large filament eruption on EK Dra \citep{Namekata22} with longer time and larger spatial scales (Section \ref{sec:result_ekdra}).

\subsection{EK Draconis} \label{sec:result_ekdra}

\subsubsection{Hydrodynamics}
\begin{figure*}[p]
\plotone{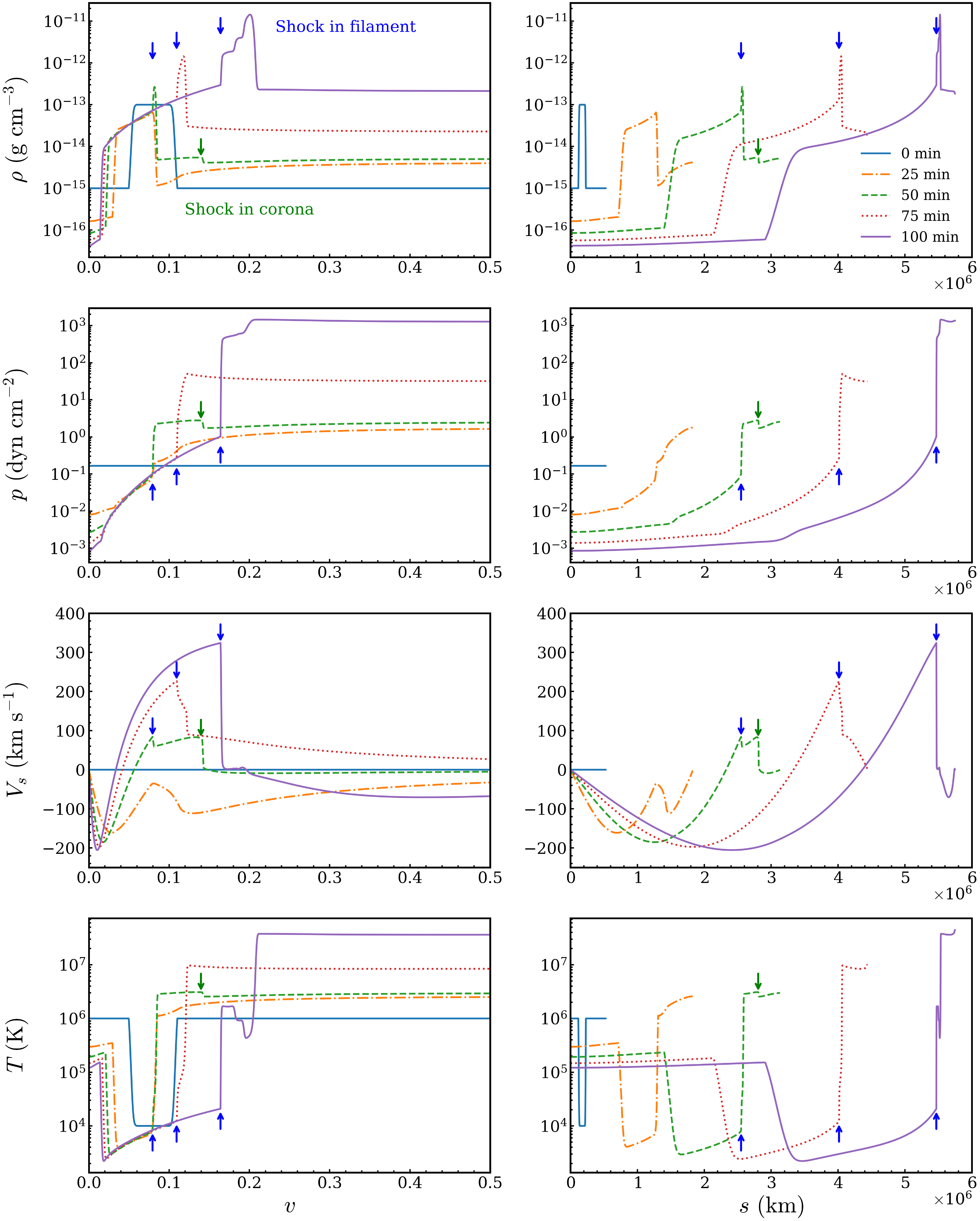} 
\caption{The temporal variation of density $\rho$, pressure $p$, velocity $V_s$, and temperature $T$, with the coordinate $v$ (left) and $s$ (right) for EK Dra at the time of $t=$ 0, 25, 50, 75, and 100 (min). 
The shock wave generated inside the filament is indicated for $t=$ 50, 75, and 100 (min) (blue arrow). A shock wave is also generated in the coronal plasma just below the filament only for $t=$ 50 (min) (green arrow).
The range of $0 \leq v \leq 0.5$ is only exhibited to enlarge the part of the filament eruption for clarity. The length from the loop top $s$ can be transformed from the coordinate $(v,u)$ with Equation \ref{eq:scale}. \label{fig:fluid_ekdra}}
\end{figure*}

Figure \ref{fig:fluid_ekdra} shows the results of the hydrodynamic simulations of the flow along the expanding loop for EK Dra (the filament eruption associated with a superflare on 5 April 2020): the temporal variations of density $\rho$, pressure $p$, velocity $V_s$, and temperature $T$, as a function of the coordinate $v$ ($\leq 0.5$) and $s$ at the time of $t$ = 0, 25, 50, 75, and 100 (min).
The characteristics of the flow in the expanding loop with the erupting filament are quite similar to those for the Sun (Figure \ref{fig:fluid_sun}).
The cool and dense filament also appears to be roughly in free-fall motion due to the gravity toward the loop bottom (rightward in each panel of Figure \ref{fig:fluid_ekdra}). As discussed in section \ref{sec:hydrosun}, the apparent large pressure imbalance for $t$ = 25 (min) is a result of adiabatic expansion of the plasma at the loop top and filament. It can be seen that a shock wave is generated inside the cool filament because the falling velocity (=84, 227, and 324 km s$^{-1}$ at the shock) exceeds the local sound speed (=15, 18, and 24 km s$^{-1}$) for $t$ = 50, 75, and 100 (min).
Only difference is that the time and spatial scales are much longer and larger than those for the Sun because the filament eruption on EK Dra was associated with a superflare (Section \ref{sec:obs}). Because of the same reason as the Sun but for larger spatial scale, a shock wave is also generated even in the coronal plasma just below the falling filament at the time of $t=50$ (min) since the falling velocity (= 80 km s$^{-1}$) is not necessarily much smaller than the local sound speed in the corona (= 290 km s$^{-1}$).

\subsubsection{Comparison with observed H$\alpha$ spectrum} \label{sec:ekdra_model}
\begin{figure*}[p]   
\plotone{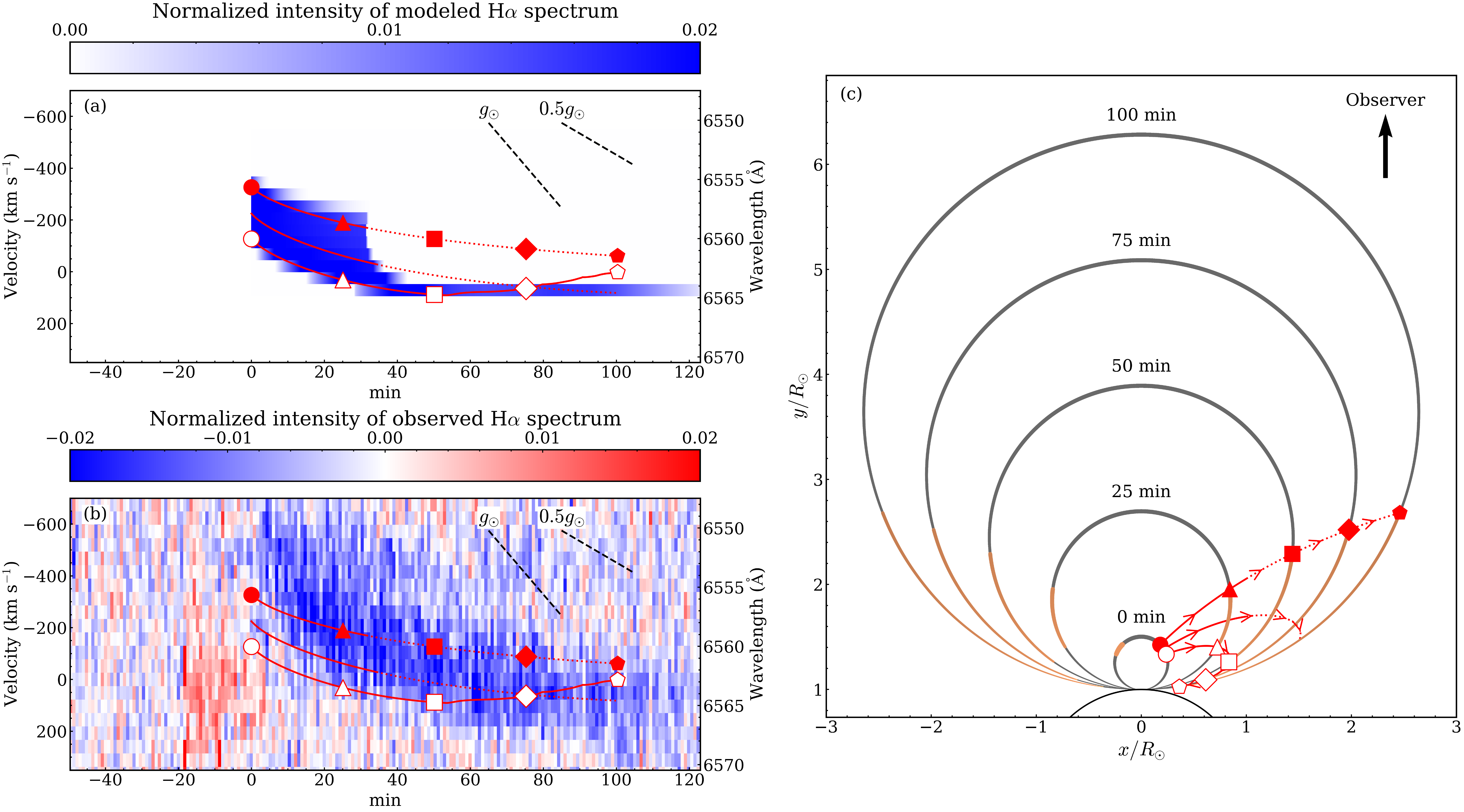} 
\caption{(a) Normalized intensity of the modeled H$\alpha$ spectrum (Section \ref{sec:vobs}), (b) normalized intensity of the observed H$\alpha$ spectrum of the filament eruption associated with a superflare on 5 April 2020 \citep[Fig.1 in][]{Namekata22}, and (c) configuration of the filament eruption in the expanding loop at the time of $t=$ 0, 25, 50, 75, and 100 (min) (orange), for EK Dra from the viewing angle of $i=0^{\circ}$.
Red solid and dotted lines with each mark respectively in and out of the stellar disk show the temporal variations of $V_{\rm obs}$ in (a, b) for the upper, middle, and lower locations of the filament eruption in (c) at each time. The time is set from the start of the H$\alpha$ absorption, corresponding to 25 min later than the peak time of the superflare (BJD 2458945.2).
The solar surface gravity $g_{\sun}$ and its half $0.5g_{\sun}$ are also represented in (a, b) for comparison.
}\label{fig:contour_out}
\end{figure*}

\begin{figure*}[p]   
\plotone{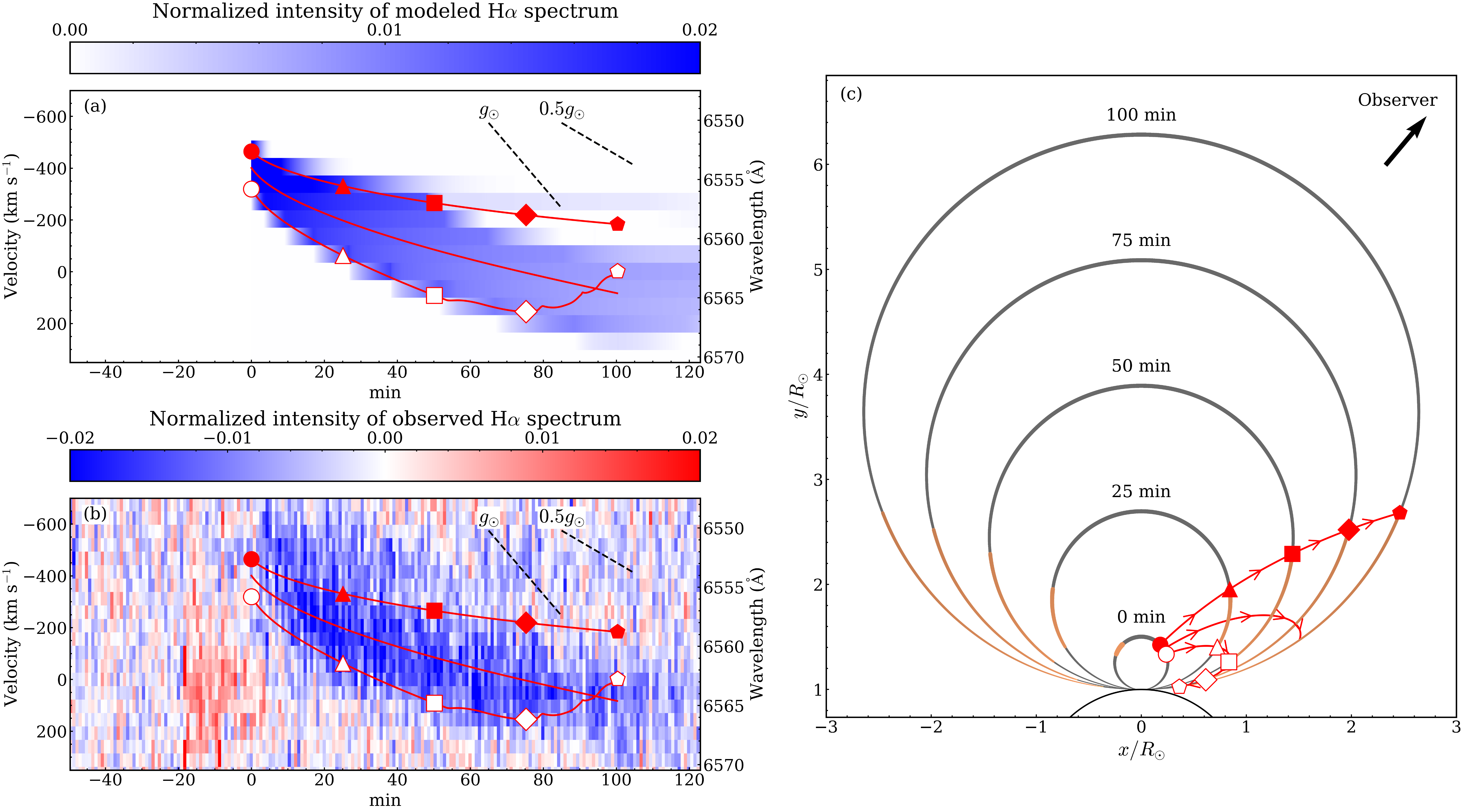} 
\caption{Same as Figure \ref{fig:contour_out}, but from the viewing angle $i=40^{\circ}$ so that the filament eruption is inside of the stellar disk to the line of sight.
} \label{fig:contour_ekdra}
\end{figure*}

\begin{figure*}[p]   
\plotone{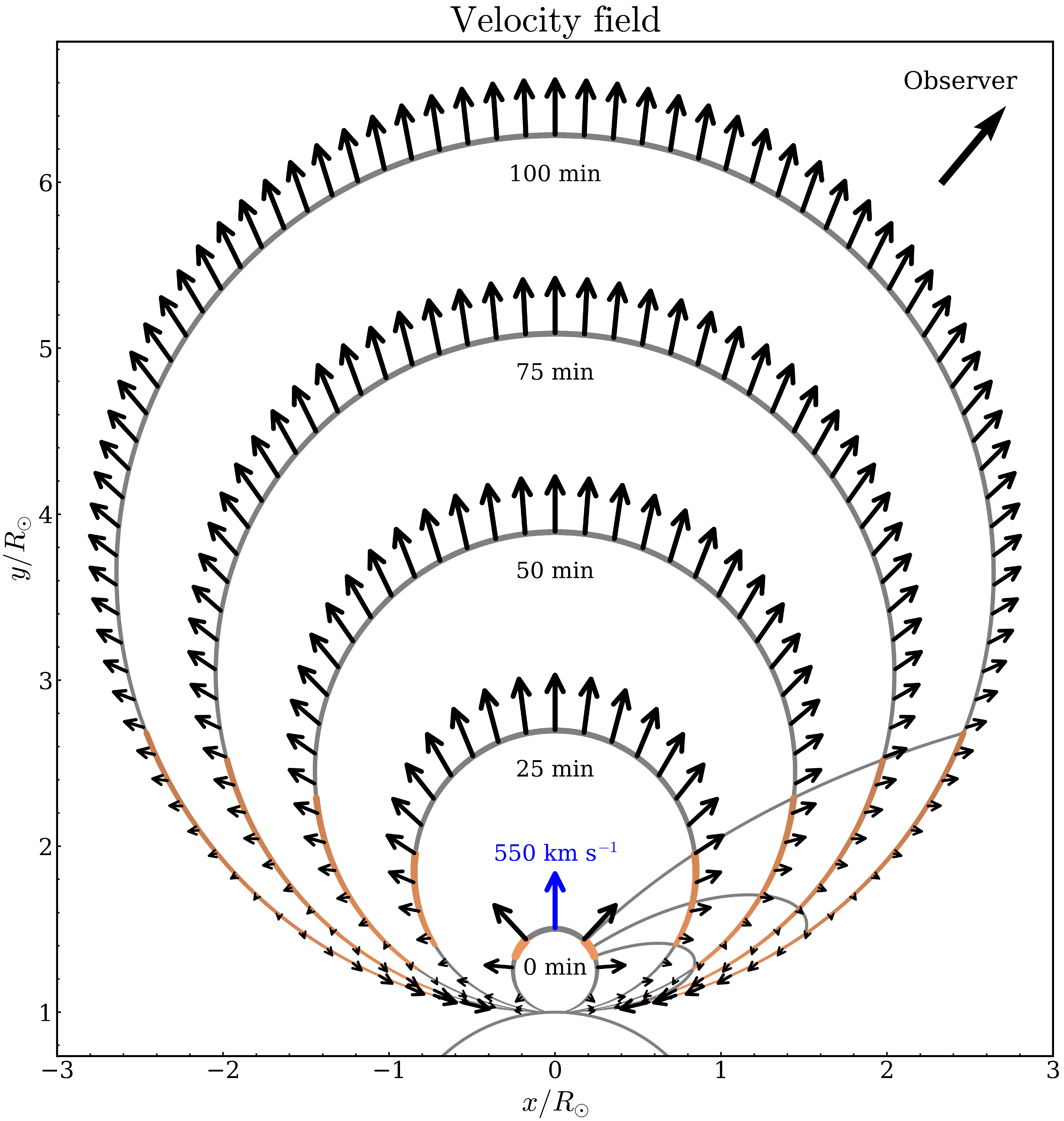} 
\caption{Velocity field $(V_x, V_y)$ of the expanding loop in the Cartesian coordinate $(x,y)$ for EK Dra (the filament eruption associated with a superflare on 5 April 2020) at the time of $t=$ 0, 25, 50, 75, and 100 (min). The filament eruption in the loop is colored in orange. The initial normal velocity of the loop top $V_{\rm top}=550$ (km s$^{-1}$) is also colored in blue for comparison of the velocity scale.}\label{fig:vector_ekdra}
\end{figure*}

Figure \ref{fig:contour_out} shows the normalized intensity of the modeled H$\alpha$ spectrum in the case of $i=0^{\circ}$ as in the case of the Sun (Figure \ref{fig:contour_sun}). However, as described in Section \ref{sec:setup}, the modeled H$\alpha$ spectrum is not reproduced from the middle time ($\sim$ 30 min) because the filament eruption is out of the stellar disk from the line of sight.
Therefore, to circumvent this problem, we set the viewing angle $i=40^{\circ}$ so that the filament eruption is observed in the stellar disk as the absorption of H$\alpha$ spectrum (Figure \ref{fig:contour_ekdra}). Then, the modeled H$\alpha$ spectrum can be reproduced with a longer time and larger spatial scale than that of the Sun.
To reproduce modeled H$\alpha$ spectrum similar to the observed H$\alpha$ spectrum, the velocity of the loop top is even larger than the escape velocity at the initial height: $V_{\rm top} = 550 > \sqrt{2 G M_\sun/(y_{\rm int }+b)} = 504$ (km s$^{-1}$).
The rarefied loop continues to expand after the filament falls to the stellar surface.
Thus, it is conceivable that hot and rarefied corona component above the loop easily exceeds the escape velocity of the star $ V_{\rm esc} = \sqrt{2 G M_\sun/r} =  618 - 276$ (km s$^{-1}$) for $r= 1 - 5 R_{\sun}$ during the acceleration of the loop and eventually propagate into interplanetary space as stellar CMEs.
Figure \ref{fig:vector_ekdra} shows the global picture for our model of the filament eruption and the expanding loop with hot and rarefied coronal plasma as a result of velocity fields along the expanding loop at the time of $t$ = 0, 25, 50, 75, and 100 (min) as in the case of the Sun (Figure \ref{fig:vector_sun}). 
Most of the coronal part of the expanding loop results in stellar CME as suggested from the previous solar observations and self-similar models (Section \ref{sec:solar_model}). 
Therefore, our model gives further support of the argument that the observations of H$\alpha$ absorption associated with a superflare on a solar-type star can be sufficient evidence of not only large filament eruption but large stellar CMEs 
\citep{Namekata22}.

\section{Conclusion and Future prospects} \label{sec:conclusion}
We examine the temporal variation of the H$\alpha$ spectrum with a simple model for the filament eruptions on the Sun and a solar-type star EK Dra by performing hydrodynamic simulation of the flow along an expanding magnetic loop emulating a filament eruption.
We calculate the modeled H$\alpha$ spectrum with the line-of-sight velocity of the filament eruption from the velocities along and normal to the loop and compare it with the observed H$\alpha$ spectrum for the Sun and EK Dra.
We also compare the result for the Sun with the spatially resolved H$\alpha$ images of the solar filament eruption.
The results are summarized as follows.

\begin{itemize}
    \item[(i)] The temporal variations of the H$\alpha$ spectrum for the Sun and EK Dra can be explained by our model with different time and spatial scales. The erupting filament of the model for the Sun roughly corresponds to the H$\alpha$ image of the observed solar filament eruption.
    \item[(ii)] Stellar coronal mass ejections were thought to be associated with the superflare on EK Dra because the rarefied loop with coronal plasma has larger velocity than the escape velocity even at the initial height and continues to expand after the filament fall to the surface.
\end{itemize}

As future works, our model is also applicable to various eruptive events on the Sun and solar-type stars, such as prominence eruptions and post-flare loops \citep[][]{Otsu22,Namekata23}. In addition, the relation between spot locations from photometry and flares occurrence therein has been investigated for M-dwarf flare stars \citep{Ikuta23} with the code of mapping starspots on the surface \citep{Ikuta20}. Therefore, we can extend to investigate the relation between spot locations and filament/prominence eruptions associated with superflares for G-dwarf flare stars (K.Namekata et al., in preparation; K.Ikuta et al., in preparation).

There are several prospects of improvement in our simple model. First, all of the H$\alpha$ spectrum is not observed as the absorption under the assumption of the optically thin condition that the optical depth is less than one.
The typical optical depth is ranged from slightly less than one up to ten (optically thick) for the eruptions on the Sun \citep{Sakaue18,Namekata22}. Then, we consider the optically thick condition in Appendix \ref{sec:thick}. The intensity results in smaller values within an order \citep[e.g.,][]{Heinzel15}, but the temporal variation of the velocity in the modeled H$\alpha$ spectrum almost corresponds to that of the optically thin condition. The H$\alpha$ spectrum should be observed as intermediate values of the result from optically thin and thick conditions. 
Second, we neglect the radiative cooling/heating, thermal conduction, and other heating by nonthermal particles \citep[e.g.,][]{Glesener13} of the filament and corona in the energy equation (Equation \ref{eq:energy}). Of course, the H$\alpha$ spectrum is from a filament eruption, and the neglect of radiative cooling/heating does not much affect the dynamics, especially for the free-fall motion of the filament due to gravity (Section \ref{sec:result}). This is because the effect of the gas pressure mainly derived from the energy equation is much smaller than that of gravity (and magnetic force) in the filament/prominence eruptions.
More realistic treatment of the energy equation should be included for modeling dynamics of the filament and coronal plasma.
We also note that the formation of the H$\alpha$ spectrum is affected by the non-LTE radiative transfer \citep[e.g.,][]{Leenaarts12,Tei20} with XEUV irradiation from the hot environment in which the filament is embedded.

In the self-similar expansion model of CMEs \citep[e.g.,][]{Gibson98}, the erupting filament/prominence is situated in the core part of the CMEs, so that magnetic fields surround the filament. In this sense, we implicitly assume that such magnetic fields in a large volume surround the filament. Thus, if a filament/prominence eruption occurs, we can expect ejections of magnetic flux in a large volume, which eventually result in CMEs. However, this situation may not be necessarily applicable to all eruptions.
Instead, it is also possible that the overarching magnetic field can suppress eruptions as observed in X-class flares on the active region with large sunspots in October 2014 \citep{Thalmann15}. 
We should also note that erupting flux tubes on the Sun are often kinked with a large amount of helical twist \citep[e.g.,][]{Parenti14}. Our simple model can be incorporated with such configuration by transforming the coordinate suitable for helically twisted flux tubes (Section \ref{sec:fluid}).
Modeling these possibilities and comparison with observations would be intriguing future works especially for EK Dra.

\begin{acknowledgements}
We sincerely appreciate the referee for providing careful feedback that helped to improve both content of this manuscript and the clarity.
We appreciate Takako T. Ishii for providing data of the H$\alpha$ images of the solar filament eruption.  
We also thank Takato Otsu, Ayumi Asai, Kosuke Namekata, Yuta Notsu, Hiroyuki Maehara, and Daisaku Nogami for their observational discussion.
Numerical computations were carried out on Yukawa-21 at the Yukawa Institute for Theoretical Physics, Kyoto University.
K.I. and K.S. are supported by JSPS KAKENHI Grant Number JP21H01131. 
K.I. is also supported by JST CREST Grant Number JPMJCR1761.
\end{acknowledgements}

\appendix

\section{Optically thick condition}\label{sec:thick}
\begin{figure*}[p]
\plotone{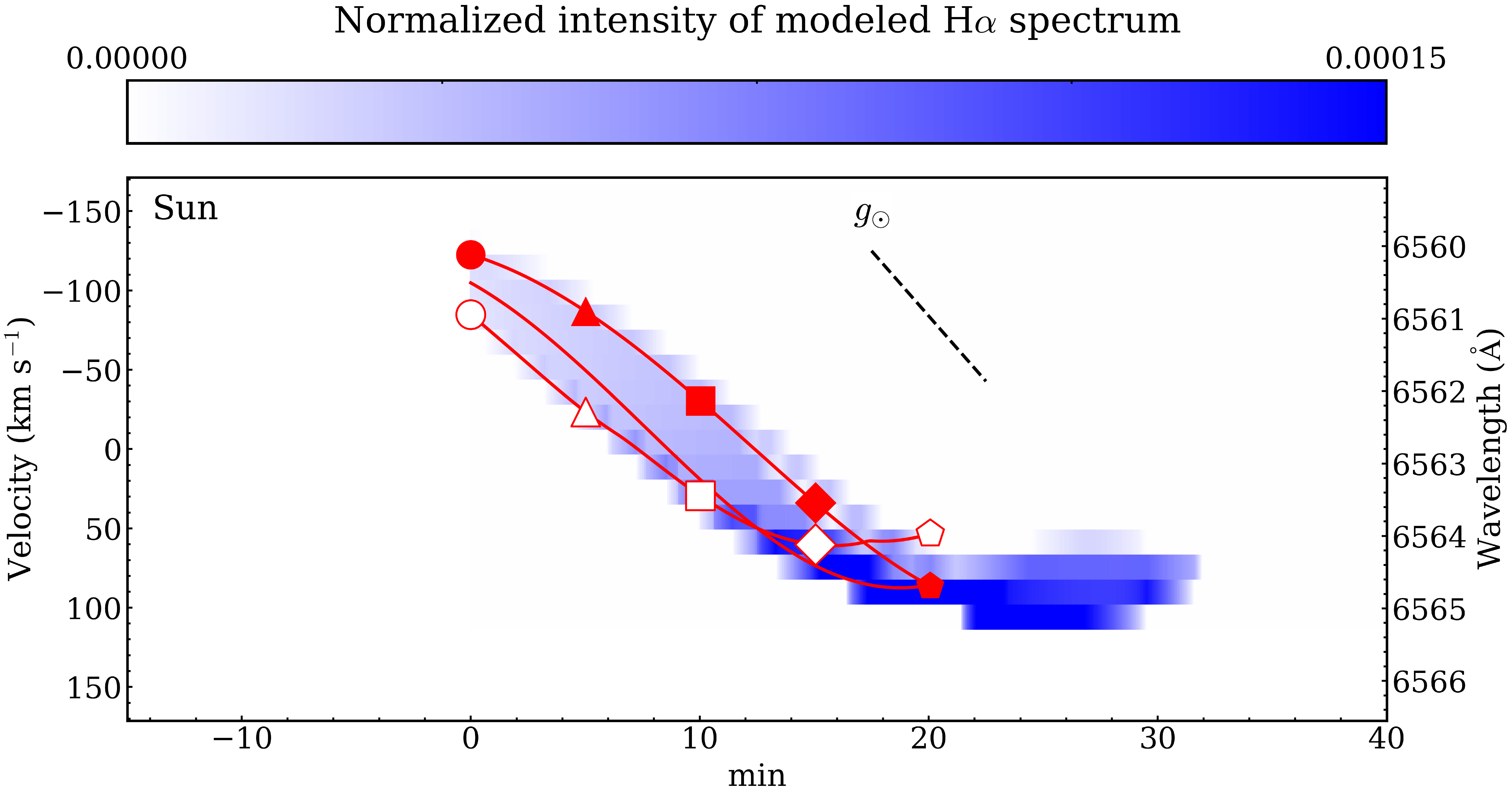}
\plotone{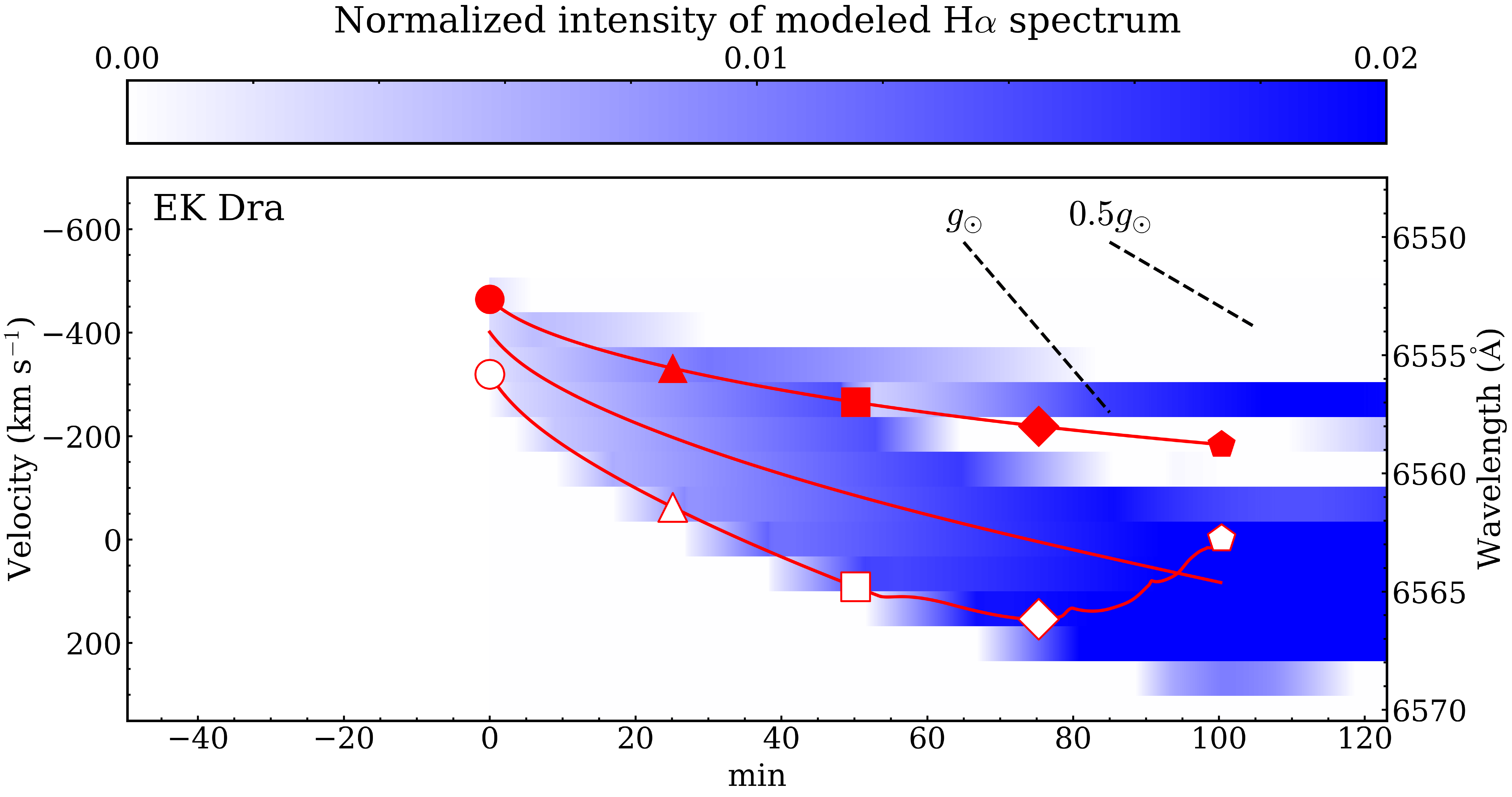}
\caption{Normalized intensity of the modeled H$\alpha$ spectrum under the optically thick condition for the Sun and EK Dra (for details, Figure \ref{fig:contour_sun} and \ref{fig:contour_ekdra}).}\label{fig:contour_thick}
\end{figure*}

For the optically thick condition, the H$\alpha$ spectrum is thought to be from the surface of filament eruptions. Thus, we recalculate the modeled H$\alpha$ spectrum only from the volume $ds_j dn_j$ (Section \ref{sec:vobs}). 
As a result shown in Figure \ref{fig:contour_thick}, for the Sun, the normalized intensity of the modeled H$\alpha$ spectrum are 0.00016, 0.00011, 0.00009, 0.00012, and 0.00014, at the time of $t$ = 0, 5, 10, 15, and 20 (min), respectively. Its value is smallest by a factor of two at the time of 10 (min).
For EK Dra, the normalized intensity of the modeled H$\alpha$ spectrum are 0.002, 0.007, 0.011, 0.013, and 0.017, at the time of $t$ = 0, 25, 50, 75, and 100 (min), respectively. Its value is smaller by an order even at the initial time.
The temporal variation of the velocity in the modeled H$\alpha$ spectrum almost corresponds to that of the optically thin condition for both of the Sun and EK Dra.
Thus, the observed H$\alpha$ spectrum is well reproduced even under the optically thick condition, and the spectrum should be observed as intermediate values of optically thin and thick conditions \citep[e.g.,][]{Heinzel15}.

\bibliography{filament}{}

\restartappendixnumbering

\bibliographystyle{aasjournal}

\end{document}